\documentclass{article}

\begin{document}

\title{Brans-Dicke theory: Jordan vs Einstein Frame}
\author{ A. Bhadra \thanks{%
Email address: aru\_bhadra@yahoo.com} \\
Admn. Block, University of North Bengal, Siliguri (WB) -743 013, India \\
K. Sarkar \thanks{%
Email address: kabita\_id@rediffmail.com}, D. P. Datta \thanks{%
Email address: dp\_datta@yahoo.com} and K. K. Nandi \thanks{%
Email address: kamalnandi1952@yahoo.co.in} \\
Department of Mathematics, University of North Bengal, Siliguri (WB) -734
013, India }
\date{}
\maketitle

\begin{abstract}
It is well known that, in contrast to general relativity, there are two
conformally related frames, the Jordan frame and the Einstein frame, in
which the Brans-Dicke theory, a prototype of generic scalar-tensor theory,
can be formulated. There is a long standing debate on the physical
equivalence of the formulations in these two different frames. It is shown
here that gravitational deflection of light to second order accuracy may
observationally distinguish the two versions of the Brans-Dicke theory.
\end{abstract}

PACS number(s): 04.50.+h \newline

\pagebreak

\section*{I. Introduction}

The Brans-Dicke (BD) theory [1], which describes gravitation through a
spacetime metric ($g_{\mu \nu }$) and a massless scalar field ($\phi $), is
a modification or rather generalization of General Relativity (GR). The
theory has recently received widespread attention due to the fact that it
arises naturally as the low energy limit of many theories of quantum gravity
such as the supersymmetric string theory or the Kaluza-Klein theory and is
also found to be consistent with present cosmological observations [2-7].

As a generic aspect of any scalar-tensor theory, two frames are available to
describe the BD theory. One\ frame is called the Jordan frame (JF) in which
the BD field equations were originally written and the BD scalar field
played the role of a spin-0 component of gravity. The other is the
conformally rescaled Einstein frame (EF) in which the scalar field plays the
role of a source matter field. There is a long standing debate as to whether
the descriptions of the BD theory in the two frames, JF and EF, should be
considered physically eqivalent. In order to get a flavor of this debate and
the resulting confusion, we should only say that physicists are divided
roughly into six groups depending on their attitude to the question. They
can be listed as follows. Some authors: (1) neglect the issue, (2) think
that the two frames are physically equivalent, (3) consider them physically
nonequivalent but do not provide supporting arguments, (4) regard only JF as
physical but, if necessary, use EF for mathematical convenience, (5) regard
only EF as physical, (6) belong to two or more of the above categories! For
a detailed account, see the review [8].

It has been argued in the literature that the physical frame is the one in
which matter couples directly (as opposed to anomalously [9]) to it,
particles have constant mass and move on geodesics of the physical metric so
that the physical stress tensor is conserved [10]. In the non-physical
frame, like the EF, particles have scalar field dependent masses and do not
move along the geodesics of the EF metric due to the occurrence of a scalar
field dependent force. This fact is manifest in the conservation of the sum
of the energy momentum tensor in the JF, the scalar field and the
cosmological term (if it is taken into account). Although, it is a matter of
theoretical interpretation which frame is the \textquotedblleft true" frame,
the physical metric is still the one that defines lengths and rates of ideal
clocks and it is the one that should be compared with observables.

Flanagan [11] has argued that all physical observables are conformal frame
invariants. Some works in cosmology do show that it is indeed the case
[12-14]. Therefore, the question arises if we can take the deflection of
light as a physical observable. We state that the deflection angle is
definitely a physical observable. In fact, the observed deflection of light
of appropriate magnitude by solar gravity provided the first experimental
proof of general relativity. The difference in the deflection angle in JF
and EF is not an effect of choosing different physical units. See the end of
Sec.IIIB for clarifications.

To resolve the issue in question (JF vs EF) in a more conclusive manner, we
feel that it is necessary to go beyond mere (mostly speculative) theoretical
arguments favoring one position or the other as listed above, and refer to a
tangible \textit{observational} ground to determine if the two frames are
physically equivalent. There exist only very few works in this direction
[15,16]. The situation is that the distinctive observational features that
emerged from these works, such as different interaction nature of
gravitational wave with gravitational detectors [15], are unlikely to be
observed experimentally in the near future. Therefore, in the present work,
we consider a more pragmatic premise, namely, the deflection of light by
gravity up to second order in gravitational strength in both versions of the
BD theory. The aim is to explore whether both formulations give same results
or not.

The plan of the paper is the following: In the next section II, we briefly
review the gravitational deflection of light in a generic static,
spherically symmetric spacetime (in isotropic coordinates). Explicit
expressions for the second order light deflection in JFBD theory and EFBD
theory are obtained in section III that also includes a discussion on the
matter of changing units. Finally, the results are discussed in section IV. 
\newline

\section*{II. Second-order deflection angle}

A general static, spherically symmetric spacetime in isotropic coordinates
is given by (geometrized units are used, unless specifically restored: $G=1$%
, $c=1$) 
\begin{equation}
ds^{2}=B(\rho )dt^{2}-A(\rho )\left( d\rho ^{2}+\rho ^{2}d\theta ^{2}+\rho
^{2}sin^{2}\theta d\varphi ^{2}\right) 
\end{equation}%
The equation of the orbital motion of test partcles can be obtained from the
geodesic equations and is given by 
\begin{equation}
\frac{d\varphi }{d\rho }=\frac{1}{\rho \sqrt{\frac{1}{Bb^{2}}-\frac{E}{b^{2}}%
-\frac{1}{\rho ^{2}A}}}
\end{equation}%
where $b=\frac{J}{E}$ is the impact parameter (the perpendicular distance
between the gravitating object and the tangent to the null geodesic) at
large distances, $E$ and $J$ are proportional to the asymptotic energy and
angular momentum of the particle. Because of the spherically symmetry, the
motion has been considered only in the equatorial plane ($\theta =\frac{\pi 
}{2}$). Following the standard treatment [17], the expression for the
deflection angle for the light rays can be written as 
\begin{equation}
\alpha (\rho _{o})=I(\rho _{o})-\pi 
\end{equation}%
where 
\begin{equation}
I(\rho _{o})=2\int_{\rho _{o}}^{\infty }\frac{d\rho }{\rho }\left[ \left( 
\frac{\rho }{\rho _{o}}\right) ^{2}\frac{A(\rho )B(\rho _{o})}{A(\rho
_{o})B(\rho )}-1\right] ^{-\frac{1}{2}}\;
\end{equation}%
$\rho _{o}$ being the distance of closest approach. The relation between the
impact parameter and the distance of closest approach follows from the
conservation of the angular momentum of the scattering process and is given
by 
\begin{equation}
b(\rho _{o})=\rho _{o}\sqrt{\frac{A(\rho _{o})}{B(\rho _{o})}}
\end{equation}%
\ \ \ \ \ Usually the Parameterized Post-Newtonian (PPN) formalism is
employed to describe the gravitational theories in the solar system and also
to compare predictions of general relativity to the results predicted by an
alternative metric theory of gravity. This method actually is an
approximation for obtaining the dynamics of a particle (in a weak
gravitational field of a slowly moving gravitating source) to one higher
order in $\frac{M}{\rho }$ than given by the Newtonian mechanics. The
calculation of particle dynamics typically requires knowledge of $g_{oo}$
more accurately than $g_{ij}$. But as noted in [16], understanding about the
light propagation in curved spacetime to any given order needs knowledge of
every term to that order. \newline
\ \ \ Following the standard PPN expansion treatment, we assume that the
metric tensor is equal to the Minkowski tensor $\eta _{\mu \nu }$ plus
corrections in the form of expansions in powers of $\frac{M}{\rho }$ ($M$ is
the mass of the source object). Considering only up to the second-order
corrections terms, we have 
\begin{equation}
B(\rho )=1-2\frac{M}{\rho }+2\beta \frac{M^{2}}{\rho ^{2}}
\end{equation}%
\begin{equation}
A(\rho )=1+2\gamma \frac{M}{\rho }+\frac{3}{2}\delta \frac{M^{2}}{\rho ^{2}}
\end{equation}%
$\beta ,\gamma $ are the PPN parameters (also known as the Eddington
parameters), $\delta $ can be considered as the post-PPN parameter. Several
of these parameters are different for different theories. In general
relativity all of them are equal to $1$ as can be readily checked by
expanding the Schwarzschild metric. \newline
The expression for the angle of light deflection up to the second order
follows from Eq.(4) and is given by 
\begin{equation}
\alpha =2(1+\gamma )\frac{M}{\rho _{o}}+\left[ \left( 2(1+\gamma )-\beta +%
\frac{3}{4}\delta \right) \pi -2(1+\gamma )^{2}\right] \left( \frac{M}{\rho
_{o}}\right) ^{2}
\end{equation}%
It is important to note that the term representing the second order effect
contains all the three parameters $\beta $, $\gamma $, and $\delta $. So,
knowing these PPN and post PPN parameters, the second order effects on
deflection angle for any metric theory of gravity can be estimated readily
from the above expression. For the Schwarzschild metric, the deflection
angle is given by 
\begin{equation}
\alpha =4\frac{M}{\rho _{o}}+\left[ \frac{15\pi }{16}-2\right] \frac{4M^{2}}{%
\rho _{o}^{2}}
\end{equation}%
A limitation of the expression (8) is that it depends on the coordinate
variable $\rho $. However, it can also be expressed in terms of coordinate
independent variables, such as the impact parameter. In that case, the
deflection angle reduces to 
\begin{equation}
\alpha =2(1+\gamma )\frac{M}{b}+\left[ 2(1+\gamma )-\beta +\frac{3}{4}\delta %
\right] \frac{\pi M^{2}}{b^{2}}
\end{equation}

\section*{III. Deflection angle in the BD theory}

The expressions of the Eddington parameters $\beta$ and $\gamma$ for the
theories under investigation are already known. For the BD theory in the
Jordan frame (JFBD) these two PPN parameters are $\beta=1$, $\gamma=\frac{%
\omega+1}{\omega+2}$, whereas for the BD theory in the Einstein frame (EFBD)
both parameters are equal to $1$. So our main task is to calculate the post
PPN parameter $\delta$ for these theories. The parameter $\delta$ occurs
only in the metric coefficient $g_{ij}$. So it is enough for us to consider
only the static case. \newline

\subsection*{A. The JFBD theory}

The scalar field in JFBD theory acts as the source of the (local)
gravitational coupling with $G\sim \phi ^{-1}$. As a consequence, the
gravitational \textquotedblleft constant\textquotedblright\ is not in fact a
constant but is determined by the total matter in the universe through an
auxiliary scalar field equation. The scalar field couples to both matter and
spacetime geometry and the strength of the coupling is represented by a
single dimensionless constant parameter $\omega $. It is generally
considered that under the limit $\omega \rightarrow \infty $, the vacuum (or
for traceless matter field) BD theory (and its dynamic generalization)
reduces to the GR but the recent finding suggests that such a convergence is
not always true [18-22]. \newline
\ \ \ In the Jordan conformal frame, the BD action takes the form 
\begin{equation}
\mathcal{A}=\frac{1}{16\pi }\int d^{4}x\sqrt{-g}\left( \phi R-\frac{\omega }{%
\phi }g^{\mu \nu }\phi _{,\mu }\phi _{,\nu }+\mathcal{L}_{matter}\right) 
\end{equation}%
where $\mathcal{L}_{matter}$ is the Lagrangian density of ordinary matter
and $R$ is the Ricci scalar. As stated earlier, the theory is constrained by
the solar system experiments. The recent conjunction experiment with Cassini
spacecraft constrains the value of the coupling constant as $|\omega
|>5\times 10^{4}$ [23]. \newline
\ \ \ The static spherically symmetric matter free solution of the BD theory
in isotropic coordinates is given by [24,25]: 
\begin{equation}
ds^{2}=+\left( \frac{1-\frac{B}{\rho }}{1+\frac{B}{\rho }}\right) ^{\frac{2}{%
\lambda }}dt^{2}-\left( 1+\frac{B}{\rho }\right) ^{4}\left( \frac{1-\frac{B}{%
\rho }}{1+\frac{B}{\rho }}\right) ^{\frac{2(\lambda -C-1)}{\lambda }}\left(
d\rho ^{2}+\rho ^{2}d\theta ^{2}+\rho ^{2}sin^{2}\theta d\phi ^{2}\right) 
\end{equation}%
\begin{equation}
\phi =\phi _{0}\left( \frac{1-\frac{B}{\rho }}{1+\frac{B}{\rho }}\right) ^{%
\frac{C}{\lambda }}
\end{equation}%
with 
\begin{equation}
\lambda ^{2}=(C+1)^{2}-C\left( 1-\frac{\omega C}{2}\right) 
\end{equation}%
where $B,C$ are constants of integration. By the weak field Newtonian
approximation, we can set $\frac{4B}{\lambda }=2GM$, where $G$ is the
gravitational constant measured by a Cavendish or a similar experiment and $M
$ is the gravitating mass. Further, by matching the interior and exterior
(due to physically reasonable spherically symmetric matter source) scalar
fields, the constant $C$ can be identified as $C=\frac{1}{\omega +2}$[17].
These are the standard procedues for fixing constants of a metric theory of
gravity. Expanding the metric coefficients and retaining only up to the
second order terms in $\frac{M}{\rho }$, we get the parameter $\delta $ as 
\begin{equation}
\delta =1-\frac{15\omega +22}{6(\omega +2)^{2}}
\end{equation}%
Hence, finally, the deflection angle becomes 
\begin{equation}
\alpha =\left( \frac{2\omega +3}{2\omega +4}\right) \frac{4M}{\rho _{o}}+%
\left[ \left( \frac{2\omega +3}{2\omega +4}-\frac{15\omega +22}{8(2\omega
+4)^{2}}-\frac{1}{16}\right) \pi -2\left( \frac{2\omega +3}{2\omega +4}%
\right) ^{2}\right] \frac{4M^{2}}{\rho _{o}^{2}}
\end{equation}%
In the limit $\omega \rightarrow \infty $, the above expression reduces to
the general relativity value. The deflection angle can also be readily
expressed in terms of impact parameter using Eqs. (10) and (15).

\subsection*{B. The EFBD theory}

Recent cosmological observations indicate that the universe is undergoing
cosmic acceleration and is dominated by a dark energy component with
negative pressure [26-29]. Cosmological constant ($\Lambda $) is a
straightforward and natural candidate for such a component. However, the
observational upper limit on $\Lambda $ is more than $120$ orders smaller
than what is expected naturally from a vacuum energy originating at the
Planck time. An alternative realization of dark energy is in the form of a
minimally coupled scalar field $\phi $ with a specific potential $U(\phi )$
(the so called `quintessence') whose slowly varying energy density would
mimic an effective cosmological constant. This is very reminiscent of the
mechanism producing the inflationary phase. A minimally coupled scalar field
is, thus, an attractive possibility in modern cosmology. \newline
The action for the EFBD theory is 
\begin{equation}
\mathcal{A}=\int \sqrt{-\tilde{g}}d^{4}x\left( \tilde{R}+\mu \tilde{g}%
^{\alpha \beta }\tilde{\phi}_{,\alpha }\tilde{\phi}_{,\beta }\right) 
\end{equation}%
This action is obtained from the action (11) by conformal transformation of
the metric $\tilde{g}_{\alpha \beta }=\phi g_{\alpha \beta }$ and a
redefinition of the scalar $\widetilde{\phi }=\left( \frac{2\omega +3}{2\mu }%
\right) ^{1/2}\ln \phi $. The extra constant $\mu $ is introduced here to
fix the sign of the kinetic term, but it does not appear in metric
observations.

A static spherically symmetric vacuum solution to the EFBD theory (with the
cosmological constant $\Lambda =0$) is the well known Buchdahl solution [30]
which is also variously referred (as demonstrated in [31]) to as JNW [32] or
Wyman solution [33]. Like its counterpart (Schwarzschild solution) in GR,
this solution also correctly explains all the post-Newtonian tests of GR.
However, in contrast to the Schwarzschild solution, Buchdahl solution does
not represent a black hole spacetime but possesses a strong globally naked
singularity, respecting the \textquotedblleft scalar no hair
theorem\textquotedblright\ [34] which purports to exclude the availability
of any knowledge of a scalar field from the exterior of a spherically
symmetric black hole. Whether a naked singularity occurs generically in a
physically realistic collapse is a subject of considerable debate [35]. 
\newline
\ \ \ The Buchdahl solution [30], in isotropic coordinates, is given by 
\begin{equation}
ds^{2}=\left( \frac{1-\frac{m}{2\rho }}{1+\frac{m}{2\rho }}\right) ^{2\xi
}dt^{2}-\left( 1-\frac{m}{2\rho }\right) ^{2(1-\xi )}\left( 1+\frac{m}{2\rho 
}\right) ^{2(1+\xi )}[d\rho ^{2}+\rho ^{2}(d\theta ^{2}+sin^{2}\theta d\phi
^{2})],
\end{equation}%
and the expression for the scalar field is given by 
\begin{equation}
\phi (\rho )=\sqrt{\frac{2(1-\xi ^{2})}{\mu }}ln\left( \frac{1-\frac{m}{%
2\rho }}{1+\frac{m}{2\rho }}\right) ,\;\rho >m/2
\end{equation}%
The Arnowitt-Deser-Misner (ADM) mass of the source corresponding to the
above solution is given by $M=\xi m$. The effect of scalar field is usually
described in terms of a scalar charge defined as $q=m\sqrt{\frac{2(1-\xi
^{2})}{\mu }}$. Expanding the metric coefficients in $\frac{M}{\rho }$ and
comparing with Eqs.(6) and (7), we get 
\begin{equation}
\delta =1-\frac{1}{3}(1-\xi ^{2})
\end{equation}%
Thus the final expression for the second-order deflection angle becomes 
\begin{equation}
\alpha =4\frac{M}{\rho _{o}}+\left[ \frac{1}{16}(14+\xi ^{2})\pi -2\right] 
\frac{4M^{2}}{\rho _{o}^{2}}
\end{equation}%
One can obtain the general relativity result by taking $\xi =1$.

Usually, a conformal transformation is regarded as a change in physical
units. Hence, a natural question is whether the difference between the
deflection angles in JF and EF, as revealed from Eqs.(16) and (21)
respectively, is an effect of selecting different conventions of physical
units. To see that this is not the case here, we get from (12), with $\frac{%
4B}{\lambda }=2GM$, to first order, $g_{tt}^{JF}=1-\frac{2GM}{\rho }$ and $%
g_{\rho \rho }^{JF}=1+\frac{2(C+2)GM}{\rho }$. Now, via the conformal
transformation $g_{\alpha \beta }^{EF}=\phi g_{\alpha \beta }^{JF}$, we get,
to first order, $g_{tt}^{EF}=1-\frac{2B(C+2)}{\lambda \rho }$ and $g_{\rho
\rho }^{EF}=1+\frac{2B(C+2)}{\lambda \rho }$. If we think that EF is the
physical frame, then, again by standard Newtonian identification, $\frac{%
2B(C+2)}{\lambda }=2GM$, we get $g_{tt}^{EF}=1-\frac{2GM}{\rho }$, $g_{\rho
\rho }^{EF}=1+\frac{2GM}{\rho }$. [Using the relation $\xi =\frac{1}{\lambda 
}\left( \frac{C+2}{2}\right) $ and $B=\frac{mG}{2}$, we do get $M=\xi m$].
Clearly, just by changing units, the components of the metric tensors in JF
and EF can not be reconciled even in the first order. One of the underlying
reasons could be that the numerical values of scalar invariants, like the
Ricci scalar, change under conformal transformations. Another reason could
be that the conformal transformation from JF to EF and its inverse do not
preserve the exact specific form of either action.

\section*{IV. Discussion}

Our main observations are as follows:\newline
\ \ \ \ \ (1) The JFBD theory contains an adjustable coupling parameter $%
\omega $. As $\omega $ increases, the post-Newtonian expansions of the BD
theory increasingly approach the corresponding GR expressions. As a result,
observations can not rule out the JFBD theory in favor of GR, but can only
place limits on the coupling parameter $\omega $. Using the present lower
bound on $\omega $ as obtained from the recent conjunction experiment with
Cassini spacecraft, we found that the second order deflection angles of
light in the GR and in the JFBD theory are \textit{same} up to an accuracy
of $100$ $pico$ $arc$ $seconds$. Hence the proposed experiments for
measuring the deflection of light to second order accuracy, such as the
LATOR experiment [36,37] which is expected to achieve an accuracy of nearly $%
10$ $nano$ $arc$ $second$ in angular measurement, would not impose any
further constraint on the coupling parameter $\omega $. However, it should
be noted that, from the accurate observation of the first order deflection
of light, the LATOR mission will measure the PPN parameter $\gamma $ very
precisely which in turn will provide better information on the value of $%
\omega $. Similar conclusions will hold also for the generalized
scalar-tensor theories for which $\omega =\omega (\phi )$. \newline
\ \ \ \ \ (2) The EFBD theory contains a scalar field that couples minimally
to gravity. Then, at the first PPN order, there is no effect of the scalar
field in the deflection angle. The difference from GR occurs only in the
magnitudes of the second and higher orders. We observe that the second order
deflection angle depends on scalar charge in addition to the ADM mass of the
source object and the bending is reduced under the effect of the scalar
charge. If the scalar charge is just $10\%$ of the total ADM mass of the
sun, the \textit{difference }between the deflection angle (up to second
order) of light in the Schwarzschild and in the Buchdahl spacetime is around 
$7$ $nano$ $arc$ $sec$. So the LATOR mission, for the first time, might
detect signatures of the minimally coupled scalar field. The difference is
significant compared to the much lesser (in principle zero, as $\omega $ is
increased without limit) difference between the Schwarzschild and JFBD
theory and its measurement could observationally distinguish between JF and
EF. This is what we wanted to argue in this paper. \\
3) We have not touched upon the cosmological issues. It might be
interesting to know if tests at the solar system level be used to set the
the boundary conditions for a cosmological problem. \\

We thank an anonymous referee for very useful comments and suggestions.

\end{document}